\documentclass[twocolumn]{elsart3}

\usepackage{amsmath,amssymb,graphicx}
\usepackage[latin1]{inputenc}

\begin{document}

\begin{frontmatter}

 \title{Point-Contact Spectroscopy on RuSr$_2$GdCu$_2$O$_8$}
 \author[unisa]{S. Piano},
  \ead{samanta@sa.infn.it}
 \author[unisa]{F. Bobba},
 \author[unisa]{F. Giubileo},
 \author[unisa]{A. Vecchione}, and
 \author[unisa]{A. M. Cucolo}
\corauth[cor1]{Corresponding author.}
 \address[unisa]{Physics Department and INFM/CNR Supermat
 Laboratory;
University of Salerno, via S. Allende, I-84081, Baronissi (SA),
Italy}

\begin{abstract}
We present Point-Contact experiments on polycrystalline
RuSr$_2$GdCu$_2$O$_8$ samples.  The majority of tunneling curves
shows a zero-bias conductance peak, which is modeled by assuming a
d-wave pairing symmetry of the superconducting order parameter.The
magnetic field dependence of the conductance spectra has been
measured in very stable junctions. In some cases, due to the
granularity of the samples, clusters of grains in series introduce
peculiar features in the conductance spectra.
\end{abstract}

%
\end{frontmatter}

\section{Introduction}\label{intro}

The symmetry of the order parameter gives important information to
understand the mechanism of the superconductivity in high-$T_c$
superconductors. A standard technique to investigate the pairing
symmetry and the presence of nodes in the superconducting energy gap
is the Point-Contact Andreev Reflection Spectroscopy (PCAR), which
consists in establishing a contact between a tip of a normal
 metal and a superconducting sample (N--S junction).

Recently a new compound of the cuprate family, the
RuSr$_2$GdCu$_2$O$_8$ (Ru-1212) \cite{Rutenato}, has drawn great
attention among theorists and experimentalists in the field of solid
state physics. Indeed, the Ru-1212 compound presents a similar
structure to YBa$_2$Cu$_3$O$_7$, with the substitution of Cu-O
chains by magnetic Ru-O$_2$ planes, so that coexistence of
superconductivity and magnetic ordering is present. The $T_c$ of
Ru-1212 depends strongly on the preparation conditions \cite{Tallon}
with some reports showing transition onset as high as $50K$
\cite{Bernhard}. The rutheno-cuprate materials also show magnetic
order around $135 K$. The magnetic order of the Ru moments seems to
be predominantly antiferromagnetic along the \textsl{c}-axis
\cite{Lynn}, while a ferromagnetic component has been observed in
plane \cite{Bernhard2}.

In this paper we report PCAR results obtained on Ru-1212 synthesized
pellets. In the majority of the cases, the conductance curves show a
triangular peak structure at zero bias. We will show that the
experimental data are well fitted by a modified \emph{d-wave} BTK
model.

\section{Experiments and theoretical modeling}\label{sample}
\begin{figure}[t!]
\centering
  \includegraphics[width=7cm]{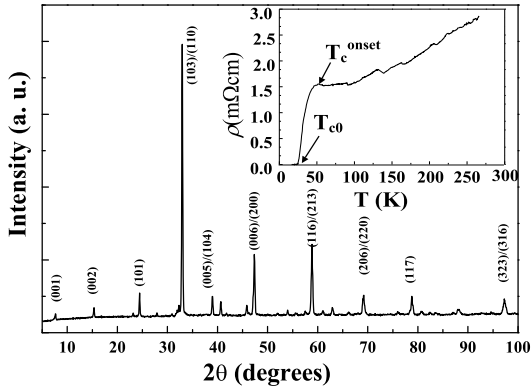}\\
  \caption{X-Ray Diffraction spectrum on Ru-1212.
  In the inset the resistivity measurement as a function of
  the temperature is shown ($T_c^{on}=43$ K, $T_ {c0}=24\,$K). }\label{raggix}
\end{figure}
 Our Ru-1212 samples were
directionally solidified pellets, prepared starting from a Ru-1212
and Ru-1210 (RuSr$_2$GdO$_6$) powders mixture (in a Ru-1212/ Ru-1210
= 0.2) by means of Top-Seeded Melt-Textured. The details of the
preparation procedure are reported in Ref. \cite{Attanasio}. X-Ray
diffraction patterns were recorded at room temperature using a
Philips PW-1700 diffractometer equipped with a Ni-filter for
CuK$\beta$ radiation. The diffraction patterns were collected over
$2\theta$ range 5°-100° with a step of 0.05° and a time per step of
20s. A typical diffraction pattern for a Ru-1212 sample is shown in
Fig. \ref{raggix}. The only phase detectable is Ru-1212 (the main
reflections are indexed) and the sample is polycrystalline.
Zero-field resistivity is shown in the inset of Fig. \ref{raggix}.
The sample is metallic down to $90 \,$ K where a saturation in the
resistivity starts to appear. This persists down to $49 \,$ K where
a weak increase of the resistivity is present at temperatures just
above the onset of the superconducting transition ($T_c^{on}=43$ K).
The sample exhibits zero resistivity at a temperature $T_ {c0}=24\,$
K.

\begin{figure}[t!]
\centering
  \includegraphics[width=7cm]{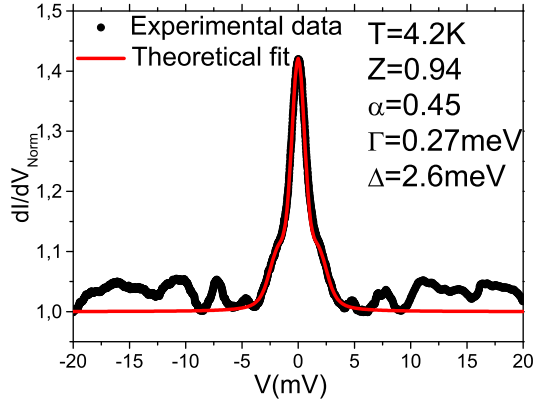}\\
  \caption{Conductance curve measured on Ru-1212/Pt-Ir point-contact junction at T=4.2K (points).
  This spectrum is fitted by BTK model with \emph{d-wave} symmetry of the order parameter (solid line).
  The fitting parameters
 are: the superconducting gap, $\Delta$, the barrier strength
 $Z$, the angle $\alpha$ of the order parameter and the smearing factor $\Gamma$.}\label{RUV1H}
\end{figure}

 To realize our
 experiments we used a
Pt-Ir tip, chemically etched in a $37\%$ solution of HCl and in an
ultrasound bath.
  Current-voltage
 ($I-V$) characteristics were obtained and a standard
 lock-in technique was used to measure the differential conductance
 ($dI/dV-V$) spectra.
Different kinds of conductance curves have been observed, majority
of which exhibiting a triangular Zero Bias Conductance Peak (ZBCP)
\cite{Samanta} with an energy width of about $10mV$, as reported in
Fig. \ref{RUV1H}.

 The triangular structure of the ZBCP can be explored in term of
 PCAR
 Spectroscopy, however there is no possibility to model this peculiar shape by
 assuming an \emph{s-wave} symmetry of the superconducting order
 parameter. Indeed we have assumed
 a \emph{d-wave} symmetry
 of the superconducting gap in a BTK modified model \cite{BTK,Tanaka}. In a \emph{d-wave}
 superconductor, the electron-like and
 hole-like quasiparticles, incident at a N--S interface,
 experience a  different sign of the order parameter, with the consequent
 formation of bound states at the Fermi energy. These states, named
 Andreev Bound States, are responsible for an increase of
 the tunneling conductance at zero-bias, in some case higher than
 2, the theoretical limit for conventional superconductors \cite{io}.

The BTK model describes the $I - V$ characteristics of a
  N--S junction separated by a
  barrier of arbitrary strength, which is modeled by a
dimensionless parameter $Z$: varying $Z$ one ranges from Andreev
Reflection regime (small $Z$) to the tunneling limit
   ($Z\gg1$).
   For an anisotropic \emph{d-wave} superconductor,
   at the given energy the tunnel current depends both on the
   incident angle $\varphi$ of the electrons at the N--S interface
   as well as on the angle $\alpha$
    of the order parameter, namely
    $\Delta_\pm=\Delta\cos[2(\alpha\mp\varphi)]$.
   It is well known that in PCAR experiments there is no preferential direction
    of the quasiparticle injection into the superconductor,
    so the tunneling current results by an integration over all directions inside a semisphere
   weighted by the scattering probability term. Moreover since our
   experiments deal with polycrystalline samples, more than one grain can be touched by
   the tip, consequently the angle $\alpha$ is a pure average fitting parameter,
   which depends on the experimental configuration.

 Fig. \ref{RUV1H} shows an experimental conductance curve at
 $T=4.2$ K, with a satisfactory best fitting curve. The used fitting parameters are: the superconducting energy gap $\Delta=2.6 \,meV$,
 the barrier strength
 $Z=0.94$, the angle $\alpha=0.45$ and the smearing factor $\Gamma=0.27 \,meV$. The factor $\Gamma$
 is a phenomenological parameter introduced to take into account pair breaking
 effects, possibly due to magnetic ordering \cite{Dynes}. We notice that the
\emph{d-wave} BTK theoretical fit correctly models the change of
slope occurring in the ZBCP at about $\pm 1 mV$.

\begin{figure}[t!] \centering
  \includegraphics[width=7cm]{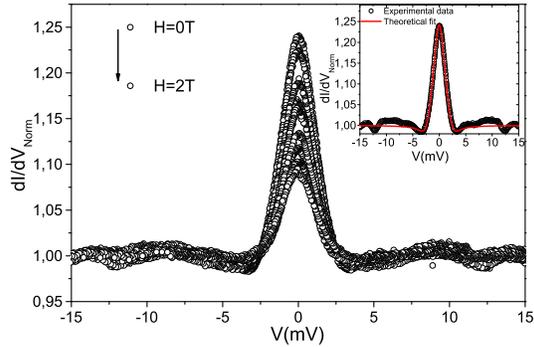}\\
  \caption{Magnetic field dependence of the $dI/dV-V$ characteristics at T=4.2K from 0 T to 2 T.
  The height and the width of the conductance peak decrease for increasing magnetic field. In the inset the zero
  field conductance spectrum is reported with the best theoretical fitting obtained by modified BTK model. The fitting
  parameters
  are: $\Delta=3.0 \, meV$, $Z=0.9$, $\alpha=0.51$ and $\Gamma=0.7 \, meV$}\label{campo}
\end{figure}

In the case of very highly stable junctions we studied the effect of
the magnetic field from zero up to $2$
 Tesla (Fig. \ref{campo}).  We observed a reduction of the height of the
 ZBCP, but a complete suppression of the
 peak and/or any splitting of this feature were not observed. This behavior is a significant evidence that the critical
 field of this sample is greater than $2$ Tesla.

 In the inset of Fig.\ref{campo} we
show the zero-field experimental data, normalized by using a
parabolic background, together with the \emph{d-wave} best fitting
curve. Values of the fitting parameters are: $\Delta=3.0 \, meV$,
 $Z=0.9$, $\alpha=0.5$ and  $\Gamma=0.6 \, meV$. We notice that both the shape of the ZBCP
 the dips around $\pm 5mV$ are correctly modeled.  Moreover, from the temperature dependence of the conductance
 curves \cite{io}, we observed
 that the ZBCP disappears at $T_c\simeq 30$ K so excluding that the formation of
 the ZBCP is due to the presence of magnetic moments at the barriers \cite{magn,cucolo}.
 Our findings about the use of the \emph{d-wave} symmetry in modeling the Ru-1212 are in
agreement with the work by G. A. Ummarino \emph{et al.}
\cite{Gonnelli}, however we infer a smaller value for the energy
gap.


Some comments are needed about the relatively low value of the
energy gap, $\Delta=(2.8 \pm 0.2) \, meV$ and the BCS ratio $2
\Delta/K_BT_C\sim 2.2$, significantly depressed in comparison with
that of conventional BCS and high $T_c$ superconductors. Moreover,
the simultaneous presence, at low temperatures, of both
ferromagnetic and superconducting orders has to be taken into
account. The Ru-1212 systems can be thought as a S/F multilayered
compound with alternating S/F layers along the \emph{c}-axis. The
two competing orders tend to mutually exclude each other, and
proximity effect in this case dramatically reduces the amplitude of
the order parameter, inducing a change of sign for critical
thickness of the magnetic layer \cite{proximity,Aprili}. In addition
to this, by means of different experimental techniques, it has been
proved that in the Ru-1212 system a large fraction of the charge
carriers is not condensed in the superconducting state at low
temperatures.

\begin{figure}[t!]
\centering
  \includegraphics[width=7cm]{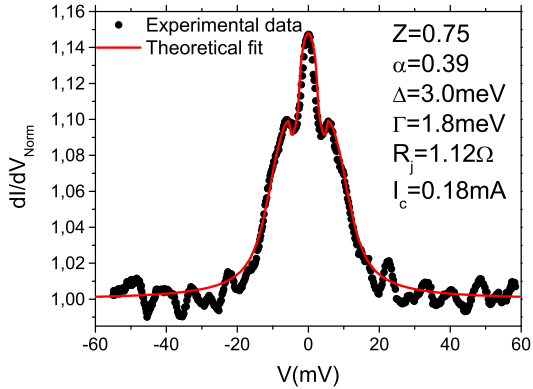}\\
  \caption{ Conductance curve measured on Ru-1212/Pt-Ir point-contact junction at T=4.2K.
  This spectrum is fitted according to our model of Josephson junction in series with the point contact,
  with a gap amplitude $\Delta=3meV$.}\label{serie}
\end{figure}

In some cases, in our experiments, we have found ZBCPs with a wider
energy amplitude, as reported in Fig.\ref{serie}.
 We speculate that the intergrain weak-coupling
 effect in polycrystalline samples plays a fundamental role in the tunneling process. Indeed a
  S--N--S or S--I--S Josephson junction can be formed in series with the N--S
 point contact one. The measured voltage is expressed as the sum of two
contributions {\cite {Lee}},
 one coming from the N--S point contact ($V_{PC}$) and the other from the intergrain
  S--N--S (S--I--S) Josephson junction ($V_J$):
 \begin{eqnarray}
V(I)&=&V_{PC}(I)+V_J(I)\;, \label{vtot}\\
\frac{dI}{dV}(V)&=&\left(\frac{dV_{PC}}{dI(V)}+\frac{dV_J}{dI(V)}\right)^{-1}\;.
  \end{eqnarray}
The $I(V)$ characteristics  are obtained by inverting
Eq.~(\ref{vtot}). We assume that the Josephson capacitance is small
and  write the Josephson voltage as follows:
\begin{eqnarray}
V_{J} &=& 0 \quad {\rm for\ } I<I_{c}\;,\nonumber \\
V_{J} &=& R_J I_{c}\sqrt{[(I/I_{c})^2-1]} \quad {\rm for\ } I>I_{c}
\;,
\end{eqnarray}
where $I_c$ and $R_J$ are the critical current and the resistance of
the Josephson junction, respectively.

The point-contact contribution to the total conductance is again
expressed in terms of the \emph{d-wave} BTK model, because an
\emph{s-wave}  symmetry, even in presence of a Josephson junction in
series, would never yield a triangular structure of the conductance
curves. In Fig. \ref{serie} we present an example of experimental
conductance with a wider ZBCP, with the corresponding theoretical
fitting according to our model of two junction in series. We observe
that, the best fitting parameters, shown in figure, are compatible
with those used for other measured junctions, and in particular the
amplitude
 of the superconducting energy gap ($\Delta=3.0 \, meV$) is consistent with
 the estimated value by our previous \emph{d-wave} fittings.

 \section{Concluding Remarks}\label{conclusion}
 We have performed Point-Contact spectroscopy experiments on Ru-1212 synthesized
 pellets and we have obtained different conductance curves,
 all characterized by  a Zero Bias Conductance Peak.
 The conductance curves with a narrow peak were selected to estimate the magnitude of order parameter.
By fitting these spectra
 we have inferred a \emph{d-wave} symmetry of the superconducting order parameter with a maximum value of the gap amplitude
$\Delta=(2.8 \pm 0.2) \, meV$.
 For the spectra with wider zero bias structures we hypothesized that the granularity of
 the samples causes the formation of a Josephson junction in series with the PC one.
 In this case, by using simplified model of a
Josephson junction in series with the N--S point contact, we were
able to fit our spectra, with a value of the order parameter
$\Delta$ in agreement with the previous results. To investigate the
interplay between superconducting and magnetic orders, we followed
the dependence of the conductance curves from an applied magnetic
field, concluding that the critical field of these samples is
greater than $2$ Tesla.

\end{document}